\documentclass[lettersize,journal]{IEEEtran}
\usepackage{amsmath,amsfonts}
\usepackage{algorithmic}
\usepackage{algorithm}
\usepackage{array}
\usepackage[caption=false,font=normalsize,labelfont=sf,textfont=sf]{subfig}
\usepackage{textcomp}
\usepackage{stfloats}
\usepackage{url}
\usepackage{verbatim}
\usepackage{graphicx}
\usepackage[nosort,nocompress]{cite}
\usepackage[T1]{fontenc}
\usepackage[utf8]{inputenc}
\usepackage{booktabs,longtable}
\newlength{\SEtablewidth}
\providecommand{\Description}[2][]{}

\begin{document}

\title{Design, development, and preliminary validity and reliability evidence of the Software Engineering Self-Efficacy Scale (SESES)}

\author{
Albert D. Ritzhaupt,
Jhonathan Sora-Cardenas,
Priya Ganapathy Prasad,
Kai A. Hackney,
Neha Rani,
\\and Pedro Guillermo Feijóo-García, ~\IEEEmembership{Member,~IEEE}%
\thanks{Albert D. Ritzhaupt is with University of Florida, Gainesville, FL, USA.}%
\thanks{Jhonathan Sora-Cardenas is with Georgia Institute of Technology, Atlanta, GA, USA.}%
\thanks{Priya Ganapathy Prasad is with University of Florida, Gainesville, FL, USA.}%
\thanks{Kai A. Hackney is with Georgia Institute of Technology, Atlanta, GA, USA.}%
\thanks{Neha Rani is with University of Florida, Gainesville, FL, USA.}%
\thanks{Pedro Guillermo Feijóo-García is with Georgia Institute of Technology, Atlanta, GA, USA.}%
\thanks{Corresponding author: Albert D. Ritzhaupt (aritzhaupt@coe.ufl.edu).}%
}

\markboth{ }%
{Ritzhaupt \MakeLowercase{\textit{et al.}}: Software Engineering Self-Efficacy Scale (SESES)}

\maketitle

\begin{abstract}
The purpose of this research is to design, develop, implement, and provide preliminary validity and reliability evidence of the \textit{Software Engineering Self-Efficacy Scale} (SESES). Framed by a conceptual framework using guidance in software engineering curriculum and concepts along with the notion of self-efficacy, we generated an initial item pool of \textit{n} = 87 items to operationalize and measure software engineering self-efficacy among undergraduate computing students. The conceptual framework traces five dimensions: 1) \textit{Requirements Engineering}, 2) \textit{Teamwork and Collaboration},\textit{ }3) \textit{Software Quality Management}, 4) \textit{Software Design and Architecture}, and 5) \textit{Software Agile Methodologies}. We pilot tested the SESES with \textit{n} = 527 undergraduate computing students who had completed a software engineering course in the current semester or a previous academic semester. We employed Exploratory Factor Analysis (EFA) with the Principal Axis Factoring method and an oblique (Promax) rotation to examine the underlying structure of the SESES, resulting in the same five internally consistent latent constructs in the conceptual framework with minimal cross-loading and a simple structure in the pattern matrix, explaining approximately 57\% of the variability in these data. Our findings suggest that software engineering self-efficacy is a multidimensional construct of five theorized and correlated, yet distinct latent factors. We unpack the limitations and delimitations of the research while exploring undergraduate computing students’ software engineering self-efficacy using necessary domain-specific measurements.
\end{abstract}

\begin{IEEEkeywords}
Software engineering, self-efficacy, Exploratory Factor Analysis, validity, reliability.
\end{IEEEkeywords}

\section{Introduction}

\IEEEPARstart{U}{ndergraduate} computing students are engaged in a multi-layered curriculum sequence that traces the foundational concepts of computing disciplinary traditions, ranging from programming fundamentals (e.g., CS1, CS2) to databases to networking to programming languages to operating systems, and several other emerging topics such as artificial intelligence, cloud computing, and data science. A long-standing hallmark of computing curricula is the swiftly evolving domain of software engineering, where computing students are exposed to terminology, concepts, frameworks, and hands-on experiences that underpin modern software development life cycles in a range of possible settings, which include large-scale, enterprise-wide software solutions that are continuously and seamlessly designed, developed, tested, and deployed to meet mission-critical objectives. Software engineering has burgeoned into its own discipline, and our long-standing computing learned societies (e.g., the Association for Computing Machinery = ACM) even provide guidelines for delivering academic programs centered on software engineering research and practice \cite{Ardis2015}. While full-fledged software engineering academic programs are available today, this current research centers on the core software engineering courses found in traditional computer science undergraduate programs (hereafter referred to as computing).

Software engineering courses in undergraduate computing programs can vary widely in learning sequence, assessments, placement within the curriculum, learning activities, and learning outcomes \cite{Hilburn2005}, depending on a number of factors, such as the needs of the local economy served by the academic program, the foundational topics used as prerequisites to the course, and the availability of software tools, faculty expertise, and learning resources \cite{UndergraduateCurricula2024}. Software engineering courses are intended to simulate real-world workplace conditions, and as such, computing educators facilitating these courses often create group projects that require students to practice teamwork and collaboration in creating and maintaining stable versions of software products \cite{Kumar2024} using modern tools (e.g., GitHub). Yet, software engineering courses are often responsible for exposing students to the full software development lifecycle (SDLC) and software methods for managing the lifecycle from start to finish, using modern development approaches such as agile methods (e.g., Scrum) and/or blended approaches (e.g., structured agile). Undergraduate computing students in a software engineering course must learn to elicit, document, verify, model, and implement users' evolving requirements and ensure the traceability of the documentation from start to finish. Further, these courses provide foundational knowledge of software analysis, design, and architecture, in which students study and apply various design patterns and principles, learn key concepts such as cohesion and coupling, and explore architectures (e.g., Model-View-Controller) for building reusable, scalable, and maintainable software artifacts\cite{Akdur2022}. These courses even span the notion of validation and verification of software, extending across the software quality management processes \cite{Ouhbi2020}. The multifaceted nature of these courses places considerable demands on students, who must simultaneously navigate technical and soft skills, including collaborative, managerial, and communication competencies, raising questions about how they perceive their own capacity to meet those demands.

For several decades, the computing educational research community has searched for ways to measure, predict, and explain educational outcomes in our disciplinary context. One promising construct widely adopted in the general educational research community is self-efficacy, which is described as a belief in one's capacity to execute the actions required to achieve specific types of performance outcomes \cite{Bandura1977}. Self-efficacy has been consistently related to several important educational outcomes in the research literature, including academic achievement \cite{Honicke2016}, perseverance \cite{Usher2019}, motivation \cite{ShenND}, and self-regulated learning \cite{Panadero2017}. For instance, self-efficacy has also been studied alongside other psychological constructs, such as the impostor phenomenon (IP) \cite{panahi2026exploratory, sora-cardenas-fie}. Though self-efficacy has been used in educational research for more than four decades, the computing educational research community has been slower to operationalize and measure this construct in domain-specific formal educational computing settings. Some domain-specific measurements of self-efficacy in computing do exist, such as computer programming self-efficacy \cite{Ramalingam1998}, computer self-efficacy \cite{Compeau1995}, computational thinking self-efficacy \cite{ComputationalThinking2019}, algorithms self-efficacy \cite{Danielsiek2018}, object-oriented programming self efficacy \cite{GanapathyPrasad2025} and even the recently established cybersecurity self-efficacy \cite{Kim2025}. Given the multilayered nature of the computing discipline, we need more domain-specific measures of self-efficacy that span multiple layers and the breadth of the computing curriculum for research, evaluation, and practice.

Bandura’s \cite{Bandura1977} account of self-efficacy includes four interrelated concepts: 1) mastery experiences (i.e., successfully performing a task), 2) vicarious experiences (i.e., observing others complete tasks successfully), 3) social persuasion (i.e., encouragement from others that one can succeed at a task), and 4) physiological and emotional states (i.e., our reactions to stress, anxiety or excitement). Each of these four dimensions can influence an individual’s self-efficacy. While these four dimensions are core components of the self-efficacy theory, educational researchers often focus on mastery experiences in measurement tools, using well-written “I can” statements to yield a domain-specific measure of self-efficacy in a targeted area of interest \cite{Bandura2006,Maurer1998}. To date, we have been unable to identify a domain-specific, valid, and reliable measurement system to operationalize and assess software engineering self-efficacy among undergraduate computing students. The need for such a measurement tool is of great practical importance to educators, practitioners, and researchers seeking to better understand undergraduate computing students' educational needs and academic performance in the context of a software engineering course that must meet increasingly insurmountable learning outcomes while simultaneously assessing a moving target. Such a measurement tool would offer a deeper exploration of the wide-range of topics covered in a software engineering course for research, evaluation, and practice. Further, such a measurement tool could be deployed by recruiters in assessing junior software engineers in early stages of onboarding. As such, we elected to use a systematic and intentional process to generate our own measure of software engineering self-efficacy.

\section{Purpose and Research Questions}

The purpose of this research study was to design, develop, implement, and provide preliminary validity and reliability evidence of a comprehensive measure of undergraduate computing students’ software engineering self-efficacy aligned to contemporary research literature and curriculum guidance on software engineering. Our guiding research questions are: 1) What are the underlying latent constructs of an undergraduate computing student’s software engineering self-efficacy? 2) What evidence of validity and reliability supports the design and use of the \textit{Software Engineering Self-Efficacy Scale} (SESES) for undergraduate computing students?, and 3) What are the differences among undergraduate computing students' demographics and backgrounds on software engineering self-efficacy? To address these guiding research questions, the present study outlines our conceptual framework for software engineering self-efficacy, our systematic design and development process, and the findings from our first administration of the SESES on a robust sample of undergraduate computing students.

\section{Review of Relevant Literature}

\subsection{Foundations of Self-Efficacy and Its Measurement in Computing}

Self-efficacy is dynamic, built primarily through mastery experiences, and developed not as a process of skill accumulation but as a generative process in which cognitive, social, emotional, and behavioral capabilities must be organized and orchestrated to serve multiple purposes \cite{Bandura1997}. Because self-efficacy is domain-specific rather than global, all-purpose measures tend to have limited predictive value when their items lack relevance to the specific domain, are cast in general terms divorced from situational demands, or generate ambiguity about what is being measured \cite{Bandura2006}. To address this, domain-specific scales must be grounded in a conceptual analysis of the domain of functioning and tailored to its specific demands \cite{Bandura2006}.

Guided by this principle of domain specificity, the computing education research community has developed a range of self-efficacy measures tailored to different areas of the discipline. A foundational measure of computer self-efficacy showed that these beliefs influence individuals' outcome expectations, emotional reactions to computers, and actual computer use \cite{Compeau1995}. Building on this work, a self-efficacy scale specific to programming in C++ was developed and validated for undergraduate students \cite{Ramalingam1998}. Beyond the development of these measures, subsequent work has examined how self-efficacy beliefs operate across different levels of specificity within computing. Initial general computer self-efficacy beliefs predict subsequent task-specific beliefs, which themselves develop over time through hands-on experience \cite{Agarwal2000}. Even so, domain-specific measures predict performance in their corresponding domains more accurately than general computer self-efficacy measures \cite{Downey2009}, a pattern later confirmed in subsequent work \cite{Davazdahemami2018}. Consistent with the limited transfer of self-efficacy beliefs across domains \cite{Ho2011}, adapting an existing programming or general computing scale would be less suitable than a domain-specific instrument for representing students' beliefs about their capabilities in software engineering.

\subsection{Self-Efficacy in the Software Engineering Classroom}

Software engineering encompasses a broad range of activities that extend beyond writing code, requiring students to navigate multiple areas of practice throughout a course. When self-efficacy has been examined in this context, the available research has addressed the construct at either a general level or within isolated components of the discipline. At the general level, self-efficacy has been shown to increase among undergraduate students engaged in problem-based learning within a software engineering capstone course \cite{Dunlap2005}, and students participating in open source software development courses reported changes in their self-efficacy beliefs as a result of engaging in large-scale software projects \cite{Salerno2023}. Beyond these broader examinations, other work has focused on specific components within the discipline. Communication self-efficacy was found to increase significantly among students in software engineering project courses after the completion of a project-based learning experience \cite{Hiranrat2023}, and debugging self-efficacy was found to decrease when undergraduate students used off-the-shelf professional tools, while simplified tools mitigated that negative effect \cite{Gilbert2024}. Taken together, this body of work shows that self-efficacy has been recognized as relevant across different areas of software engineering, yet each study has addressed only a single component or a general view of the discipline, leaving the domain without a measure that captures students' beliefs across its multiple areas of practice.

\subsection{Curricular Alignment to Software Engineering}

Moving beyond this fragmentation requires a structured view of what software engineering, as a domain, actually comprises. The Software Engineering Body of Knowledge (SWEBOK) , produced by the Institute of Electrical and Electronics Engineers’ (IEEE) Computer Society, characterizes the discipline from the standpoint of professional practice \cite{Bourque2014}. SWEBOK organizes software engineering into eighteen knowledge areas, ranging from software requirements, design, construction, and testing to software quality, configuration management, and engineering process, among others. At the level of academic preparation, the IEEE/ACM/Association for the Advancement of Artificial Intelligence (AAAI) Computer Science Curricula 2023 (CS2023) serves as guidance for undergraduate computing programs \cite{Kumar2024} . CS2023 organizes software engineering around nine knowledge areas: Teamwork, Tools and Environments, Product Requirements, Software Design, Software Construction, Software Verification and Validation, Refactoring and Code Evolution, Software Reliability, and Formal Methods. These nine areas describe what a complete software engineering program should cover. Beyond curricular and professional frames, the research literature has synthesized how software engineering competencies have been studied. A study reviewed 60 primary studies on software engineering competencies and identified 49 essential competencies organized into eleven thematic categories \cite{Assyne2022}. These include programming and coding, problem analysis, project management, teamwork, leadership, communication, customer orientation, creativity, and three categories grouped under personal traits. Read together, these three sources characterize software engineering from complementary positions: SWEBOK from the standpoint of professional practice, CS2023 from the standpoint of academic programs, and Assyne et al. from the standpoint of the accumulated research literature. The areas they identify, however, describe the full scope of the discipline rather than what any single course of software engineering education encompasses.

Across the three sources, several areas converge to constitute software engineering as a domain. Requirements engineering appears as Software Requirements in SWEBOK, as Product Requirements in CS2023, and as part of customer-oriented competencies in Assyne et al. Software design and architecture appears as Software Design and Software Architecture in SWEBOK, as Software Design in CS2023, and as problem analysis competencies in Assyne et al. Software quality management appears as Software Quality in SWEBOK and as the combination of Verification and Validation and Software Reliability in CS2023. Teamwork and collaboration appears as a recognized essential practice in SWEBOK, as Teamwork in CS2023, and as teamwork skills in Assyne et al. Agile methodologies, while not consolidated as a standalone area in any of the three sources, appear transversally across SWEBOK and CS2023 and are identified in Assyne et al. as an underrepresented area in software engineering education relative to their prominence in practice. The convergence of these areas across professional, academic, and research-based characterizations of the discipline supports their identification as the dimensions through which self-efficacy in software engineering can be meaningfully examined.

Because software engineering is a structured domain composed of distinguishable practices, beliefs about one's capability in software engineering are not a single global belief but a structured set of beliefs that correspond to those practices. A student may feel highly capable of contributing to a team while feeling less capable of designing a system, or confident in writing code while uncertain about validating it. Capturing these beliefs requires a measurement approach that reflects the multidimensional nature of the domain rather than collapsing it into a single score. The areas identified through this convergence thus provide the conceptual foundation for developing an instrument that examines self-efficacy along multiple dimensions of software engineering practice.

\section{Conceptual Framework}

Our conceptual framework is inspired by contemporary literature on software engineering research and practice, formal software engineering curriculum guidelines \cite{Ardis2015}, and framed by the concept of self-efficacy \cite{Bandura1977}, particularly the notion of mastery experiences and manifesting software engineering self-efficacy with carefully written “I can” statements \cite{Bandura2006,Maurer1998}. Figure 1 provides a visualization of our conceptual framework, which organizes the contemporary literature on software engineering and self-efficacy into five inter-related dimensions, which are used to create our initial pool of items for the SESES and guide our analyses and interpretation of our research findings. Our five dimensions of software engineering self-efficacy are:1) \textit{Requirements Engineering}, 2) \textit{Teamwork and Collaboration},\textit{ }3) \textit{Software Quality Management}, 4) \textit{Software Design and Architecture}, and 5) \textit{Software Agile Methodologies}. This section explores these dimensions and their connection to self-efficacy.

\begin{figure*}[!t]
\centering
\includegraphics[width=.74\linewidth]{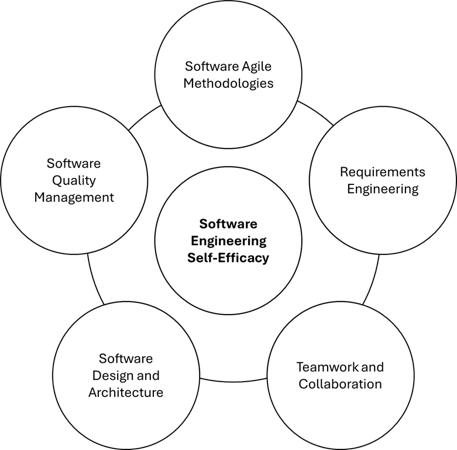}
\caption{Conceptual framework for software-engineering self-efficacy among computing undergraduate students.}
\label{fig:1}
\Description{Conceptual framework for software-engineering self-efficacy among computing undergraduate students.}
\end{figure*}

\subsection{Software Agile Methodologies}

Agile methodologies involve applying iterative principles to balance business needs with technical rigor through flexible development cycles \cite{Bourque2014}. The primary challenge for students is internalizing an ``agile mindset" rather than merely executing practices \cite{Kropp2013,Martin2017}. Within a conceptual model of software engineering, software agile methodologies self-efficacy represents a fundamental dimension of psychological adaptation to iterative uncertainty. Consequently, this domain serves as a theoretical pillar to explain how practitioners manage documenting requirements, defining scope, meeting timelines, and creating deliverables during rapid, evolving project releases.

\subsection{Software Quality Management}

Software quality management focuses on ensuring that products meet both functional and non-functional requirements through systematic constraints and process standards \cite{Bourque2014}. Although practitioners consider quality assurance a vital professional skill \cite{Lethbridge2000}, research identifies significant knowledge gaps in graduates regarding configuration and quality processes \cite{Garousi2019}. This gap highlights a critical area where software engineering self-efficacy acts as a bridge between theoretical knowledge and industrial demands. Therefore, software quality management self-efficacy is an essential component of the global software engineering construct, reflecting the belief in one's capability to maintain technical excellence.

\subsection{Requirements Engineering}

Requirements engineering is defined as the systematic process of eliciting, documenting, and validating stakeholder needs to establish a technical baseline \cite{Bourque2014}. Students often misperceive requirements engineering as a ``soft" skill, which leads to an underestimation of the analytical rigor required to resolve stakeholder conflicts \cite{Rupakheti2018}. In a conceptual framework, requirements engineering self-efficacy serves as a vital indicator of an individual’s ability to navigate early-stage project ambiguity. This domain is essential to capture the cognitive confidence necessary for translating human needs into technical specifications \cite{Daun2023}.

\subsection{Software Design and Architecture}

Software design and architecture involve the synthesis of detailed solutions to meet architecturally significant requirements while managing complex technical trade-offs \cite{Bourque2014}. The challenge in this domain lies in moving beyond basic coding to express systemic thinking, a task that often proves difficult for novice software engineers \cite{Rupakheti2015}. Within the software engineering self-efficacy construct, this area represents the psychological capacity to make high-level structural decisions under pressure. It is a cornerstone dimension that reflects the confidence to balance business goals with technical sustainability \cite{Kazman2023}.

\subsection{Teamwork and Collaboration}

Teamwork in software engineering is defined as the collaborative construction of multi-version programs, requiring technical synchronization and interdependence \cite{Kumar2024}. Unlike general collaboration, software engineering teamwork involves managing complexity and social power disparities within technical environments \cite{Moe2010,Strode2022}. Research indicates that students frequently struggle with role ambiguity and communication breakdowns during real-world projects \cite{Farre2023}. Thus, teamwork and collaboration self-efficacy constitutes a social-cognitive dimension of the software engineering construct, essential for understanding how individuals perceive their capability to function within dynamic, professional software engineering contexts.

\section{Design and Development Process}

Our approach for the design and development of the SESES followed a systematic and intentional process inspired by both Classical Test Theory \cite{Crocker1986}, contemporary accounts of software engineering in undergraduate computing curriculum \cite{Ardis2015}, and theoretical framing from self-efficacy theory \cite{Bandura1977}. We first unpacked the software engineering curriculum sequence at two public research-intensive universities in the southeastern U.S. with Accreditation Board for Engineering and Technology (ABET)-accredited computing programs by reviewing syllabi, learning resources, and assessment activities. After dissecting these courses and connecting their learning outcomes to contemporary accounts of undergraduate computing students' software engineering knowledge and skills \cite{Akdur2022}, we used the notion of self-efficacy mastery experiences \cite{Bandura1977} as a guide to create our initial domains in our conceptual framework and generate our initial item pool of 87 “I can” statements \cite{Maurer1998} connected to the conceptual framework. Each “I can” statement was inspired by the broader domain and aligned with specific accounts of undergraduate computing students’ software engineering education reflected in the coursework.

To enhance content validity of the SESES, we carefully reviewed the initial pool of items with two course instructors responsible for teaching software engineering at their respective institutions, full members of the research team with expertise in computing curriculum, software engineering, psychometrics, and educational research, and also conducted think-alouds \cite{CognitiveInterviewingND} with a small group (\textit{n} = 4) of undergraduate computing students, two from each of the two institutions involved in developing the instrument, who had successfully completed the software engineering course in prior academic semesters. We made modifications to the items throughout this process to ensure wording, concepts, and topics aligned with undergraduate computing students' software engineering self-efficacy. Finally, we subjected the final list of items to a pilot study of undergraduate computing students who had completed a software engineering course in the current semester or in previous academic semesters. This study reports on the findings of this pilot study.

\section{Method}

The present study reports the first pilot study of the SESES with a robust sample of undergraduate computing students to explore the underlying constructs of software engineering self-efficacy and to provide preliminary validity and reliability evidence for the SESES’s use. Consequently, we report the method, results, and findings of the first SESES pilot study, following the guidance of Classical Test Theory \cite{Crocker1986}.

\subsection{Educational Context}

Our sample of undergraduate computing students is drawn from three large institutions of higher education in the southeastern U.S. with ABET-accredited computing programs. Each institution placed the software engineering course at different points in an undergraduate computing student’s academic journey. Two institutions require undergraduates to have successfully completed a data structures and algorithms course before admission to the software engineering course. Another institution places the course at the senior level with prerequisites in both data structures and algorithms, and an introduction to databases courses. While each institution addresses this body of knowledge in its own way and place in the curriculum sequence, our analysis of the syllabi, learning materials, and assessments in two of these institutions revealed that the courses share many topics in common, which served as the foundation of the creation of the conceptual framework and the associated SESES item pool.

\subsection{Participants}

Our participants (\textit{n} = 527) were undergraduate computing students who had completed a software engineering course in the current semester or a previous academic semester. These student participants were recruited from three different institutions of higher education in the southeastern U.S. with ABET-accredited computing programs. Thus, there was potential for missing data across the survey items. Table 1 provides the demographic characteristics of the student participants. As shown, approximately 73\% of participants were male, 26\% were female, and approximately 1\% were Other. The vast majority of the student participants were enrolled as full-time students (97\%).

In terms of race and ethnicity, approximately 47\% of the student participants identified as Asian, 30\% identified as White/Caucasian, 8\% as Hispanic, 4\% as Black/African American, and the remaining reported Multi Ethnic (i.e., participants could select more than one) at roughly 11\%. The student participants were across the classification system with 6\% as Freshmen (first-year), 43\% as Sophomores (second-year), 33\% as Juniors (third-year), and 18\% as Seniors (fourth-year). More than 80\% of the student participants had completed at least five college computing courses, with nearly 69\% reporting having completed a computing course in high school. Approximately 32\% of the student participants had completed or were currently completing a computing internship experience. The mean age of the student participants was \textit{M} = 20.41 (\textit{SD} = 2.58), ranging from 18 to 46 years old in our sample.

\begin{table}[!t]
\caption{Demographic and background characteristics of student participants.}
\label{tab:1}
\centering
\begingroup
\small
\setlength{\tabcolsep}{4pt}
\renewcommand{\arraystretch}{1.13}
\setlength{\SEtablewidth}{\dimexpr\linewidth-4\tabcolsep\relax}
\begin{tabular}{@{}>{\raggedright\arraybackslash}p{0.76000\SEtablewidth}>{\centering\arraybackslash}p{0.12000\SEtablewidth}>{\centering\arraybackslash}p{0.12000\SEtablewidth}@{}}
\toprule
\textbf{Variable/Category} & \textbf{n} & \textbf{\%} \\
\midrule
\addlinespace[4pt]
\multicolumn{3}{@{}l@{}}{\textit{Student Gender Identity}} \\*
Male & 375 & 72.5\% \\
Female & 136 & 26.3\% \\
Other & 6 & 1.2\% \\
\addlinespace[3pt]
\addlinespace[4pt]
\multicolumn{3}{@{}l@{}}{\textit{Student Status}} \\*
Full-time student & 500 & 96.9\% \\
Part-time student & 16 & 3.1\% \\
\addlinespace[3pt]
\addlinespace[4pt]
\multicolumn{3}{@{}l@{}}{\textit{Student Race or Ethnicity}} \\*
Asian & 238 & 46.6\% \\
Black/African American & 21 & 4.1\% \\
Hispanic/Latino & 42 & 8.2\% \\
White/Caucasian & 155 & 30.3\% \\
Multi Ethnic & 55 & 10.8\% \\
\addlinespace[3pt]
\addlinespace[4pt]
\multicolumn{3}{@{}l@{}}{\textit{Student Classification}} \\*
Freshmen & 30 & 5.8\% \\
Sophomore & 222 & 43.2\% \\
Junior & 167 & 32.5\% \\
Senior & 93 & 18.1\% \\
Non-degree seeking student & 2 & 0.4\% \\
\addlinespace[3pt]
\addlinespace[4pt]
\multicolumn{3}{@{}l@{}}{\textit{Number of College Computing Courses Completed}} \\*
1 - 2 courses & 5 & 1.0\% \\
3 - 4 courses & 82 & 15.9\% \\
5 - 6 courses & 182 & 35.2\% \\
7 - 8 courses & 128 & 24.8\% \\
9 - 10 courses & 32 & 6.2\% \\
More than 10 courses & 88 & 17.0\% \\
\addlinespace[3pt]
\addlinespace[4pt]
\multicolumn{3}{@{}l@{}}{\textit{High School Computing Coursework}} \\*
Yes & 354 & 68.5\% \\
No & 163 & 31.5\% \\
\addlinespace[3pt]
\addlinespace[4pt]
\multicolumn{3}{@{}l@{}}{\textit{Current or Prior Computing Internship}} \\*
Yes & 167 & 32.3\% \\
No & 350 & 67.7\% \\
\bottomrule
\end{tabular}
\endgroup
\end{table}

\subsection{Instruments}

The primary data collection instrument was titled the \textit{Software Engineering Self-Efficacy Scale} (SESES), which included 87 “I can” self-efficacy statements organized into five domains of software engineering conceptual knowledge, as defined by contemporary software engineering literature and curriculum guidelines. The SESES uses a five-point Likert scale response format: 1) Strongly Disagree, 2) Disagree, 3) Neither Agree, Nor Disagree, 4) Agree, and 5) Strongly Agree. Participants were instructed with a short description for each domain of the SESES (e.g., \textit{Teamwork and Collaboration}) to contextualize the nature of the items in the section. In addition to the SESES, we collected basic demographic and background information (e.g., gender, race, computing background) using a short battery of items believed to be germane to the research objectives. Table 2 provides a breakdown of the five domains, the number of items within each domain, and sample items to illustrate the constructs. For a complete list of survey items, please refer to Appendix B, which includes the item-level descriptive statistics.

\begin{table*}[!t]
\caption{Software engineering domains, number of items, and sample items.}
\label{tab:2}
\centering
\begingroup
\small
\setlength{\tabcolsep}{4pt}
\renewcommand{\arraystretch}{1.13}
\setlength{\SEtablewidth}{\dimexpr\linewidth-4\tabcolsep\relax}
\begin{tabular}{@{}>{\raggedright\arraybackslash}p{0.24000\SEtablewidth}>{\centering\arraybackslash}p{0.10000\SEtablewidth}>{\raggedright\arraybackslash}p{0.66000\SEtablewidth}@{}}
\toprule
\textbf{Domain} & \textbf{Num. Items} & \textbf{Sample Items} \\
\midrule
Software Agile Methodologies & 16 & I can understand the difference between agile and waterfall methodologies.\par I can deliver measurable value for customers through my work. \\
Requirements Engineering & 19 & I can design a system architecture that satisfies the given requirements.\par I can write clean, well-documented, and maintainable code that complies with team standards. \\
Teamwork and Collaboration & 15 & I can complete my assigned responsibilities on time and communicate any delays to my team\par I can review my teammates’ code and provide constructive, respectful suggestions to improve their work \\
Software Design and Architecture & 17 & I can analyze design trade-offs (e.g., performance vs. scalability) when selecting a software architecture\par I can decompose a complex software system into components and modules that promote maintainability and reuse \\
Software Quality Management & 20 & I can recognize and address code smells or design flaws in different programming environments\par I can refactor existing code to enhance its structure and maintainability across various software systems \\
\bottomrule
\end{tabular}
\endgroup
\end{table*}

\subsection{Data Collection Procedures}

We secured Institutional Review Board (IRB) approval at the institutions in which the research team members resided. We intentionally targeted courses at three higher education institutions in the southeastern U.S. with ABET-accredited computing programs. While we focused on software engineering courses for data collection, we also targeted a handful of senior-level computing courses that required the software engineering course at the institution as a prerequisite. This approach ensured that all participants had been exposed to the SESES topics prior to completing the survey and avoided the possibility of students being double-counted. After making arrangements with a select group of computing instructors at each institution, we provided them with text for a course announcement and a link to the online survey to post in their Learning Management System (LMS) course shells. We encouraged instructors to offer a small incentive, such as extra credit, to increase participation in the research study. As such, we collected the students’ email addresses to provide to the instructors after the survey period had concluded. Thus, the survey was not anonymous but confidential. The survey was posted for three weeks in each course shell within the institutional LMS, and instructors were encouraged to mention the study during face-to-face class sessions to encourage student participation. Student participants completed the SESES survey at their own pace on their own devices in uncontrolled environments.

\subsection{Data Analysis Procedures}

We employed a multifaceted data analysis strategy to achieve the research study's scientific objectives. We employed descriptive statistics (e.g., response frequencies, means, standard deviations), internal consistency reliability, Pearson correlation analyses, and Exploratory Factor Analysis (EFA). Prior to the analyses, we examined the data for methodological assumptions (e.g., normality) and verified that the data were suitable for factor analysis.

Using EFA, we explored the optimal factor structure of the SESES instrument for undergraduate computing students, carefully removed items with problematic cross-loadings and fell short of our factor loading threshold (> 0.4), and assigned items to latent constructs using the guidance from our conceptual framework. In the EFA, we used a promax (oblique) rotation method and principal axis factoring method as we anticipated the latent factors to be correlated \cite{Grieder2022}. We used a combination of the Kaiser criterion, parallel analysis, review of the Scree plot, and the principle of parsimony \cite{Auerswald2019} (i.e., the simplest explanation and structure) to arrive at our final factor structure. We used a combination of Independent-Samples T-tests, Analysis of Variance (ANOVA) with Tukey post-hoc procedures, and non-parametric correlation analyses to examine differences in undergraduate computing students' characteristics and their software engineering self-efficacy. Given the exploratory nature of our study, we set our nominal \textit{\ensuremath{\alpha}} = .05 for all statistical tests.

\section{Results}

We organize our results by first examining the underlying factor structure of the SESES for computing undergraduates, descriptively analyzing the individual latent constructs, then exploring the relationships among these constructs, and finally, investigating the characteristics of undergraduate computing students that moderate their software engineering self-efficacy.

\subsection{Exploratory Factor Analysis}

Prior to running the EFA models, we examined the \textit{Kaiser-Meyer-Olkin} (KMO) test for sampling adequacy and the Bartlett’s test of sphericity to evaluate the suitability of the data for EFA. The KMO was calculated at 0.971, well above the 0.50 threshold for suitability for factor analysis \cite{Williams2010}. The Bartlett’s test of sphericity yielded \textit{a \ensuremath{\chi}\textsuperscript{2} = }39,299.56 (\textit{p} < .001), suggesting significant correlations among the variables in these data \cite{Williams2010}. Additionally, our item-to-participant ratio was substantial at approximately 1:6 which is below the recommendations by Kaiser criterion of achieving a 1:10 ratio for sufficient power and factor stability, but is above other thresholds suggesting the model would still be a stable measurement for the target population \cite{Morrison2009}. With these considerations in mind, we opted to continue with the EFA of these data. We ran a range of EFA models with and without restrictions on the number of factors to retain. The unconstrained model, based on the Kaiser criterion (Eigenvalue > 1.0), suggested the model could include up to 10 latent constructs. However, the 10-factor model yielded several cross-loadings among the factors, weak factor loadings below the 0.40 threshold used for the study, and overall, it lacked a simple, interpretable factor structure aligned to our conceptual framing. Figure 2 shows the Scree plot for these data with a line at the cutscore of an Eigenvalue of 1.0. One can visually inspect the diagram for an elbow-like curve when approaching the range between five and ten factors, which suggests a stable model can be found within this range of values.

After iteratively testing several EFA models by constraining the number of factors produced and studying the factor loadings and structure of the pattern matrices, we arrived at a five-factor model that explained approximately 57.2\% of the data's variability and yielded only two troubling cross-loadings across the 87 items in the initial pool. We elected to remove the two items from the analyses that showed cross-loadings (Items 28 and 40), and additionally, two items (Items 13 and 49) did not meet the factor loading threshold ( > 0.4), which resulted in them naturally dropping out of the final item pool. Our final five-factor solution retained 83 items and yielded five internally consistent latent constructs with simple structures in our pattern matrix and zero cross-loadings among the five latent factors, as shown in Table 3. The average factor loading across items and factors was approximately 0.64, which is above the minimal thresholds to secure factor stability, but still below the levels (< 0.7) deemed best for creating robust, psychometrically sound measurement tools \cite{Clark2018}. It is notable that the design inspired by our conceptual framework did not fully align, with some items and factor loadings not exactly matching the a priori prescribed structure; yet the underlying essence of the latent constructs remained relatively stable, and the remaining items meaningfully loaded on the latent factors that still manifested the underlying theory behind the five-dimensions noted in the conceptual framework. To reference the factor loadings in the pattern matrix, please see Appendix A or you can simply visually scan Figure 3, which provides a path model of the five-factor solution.

\begin{figure*}[!t]
\centering
\includegraphics[width=.95\linewidth]{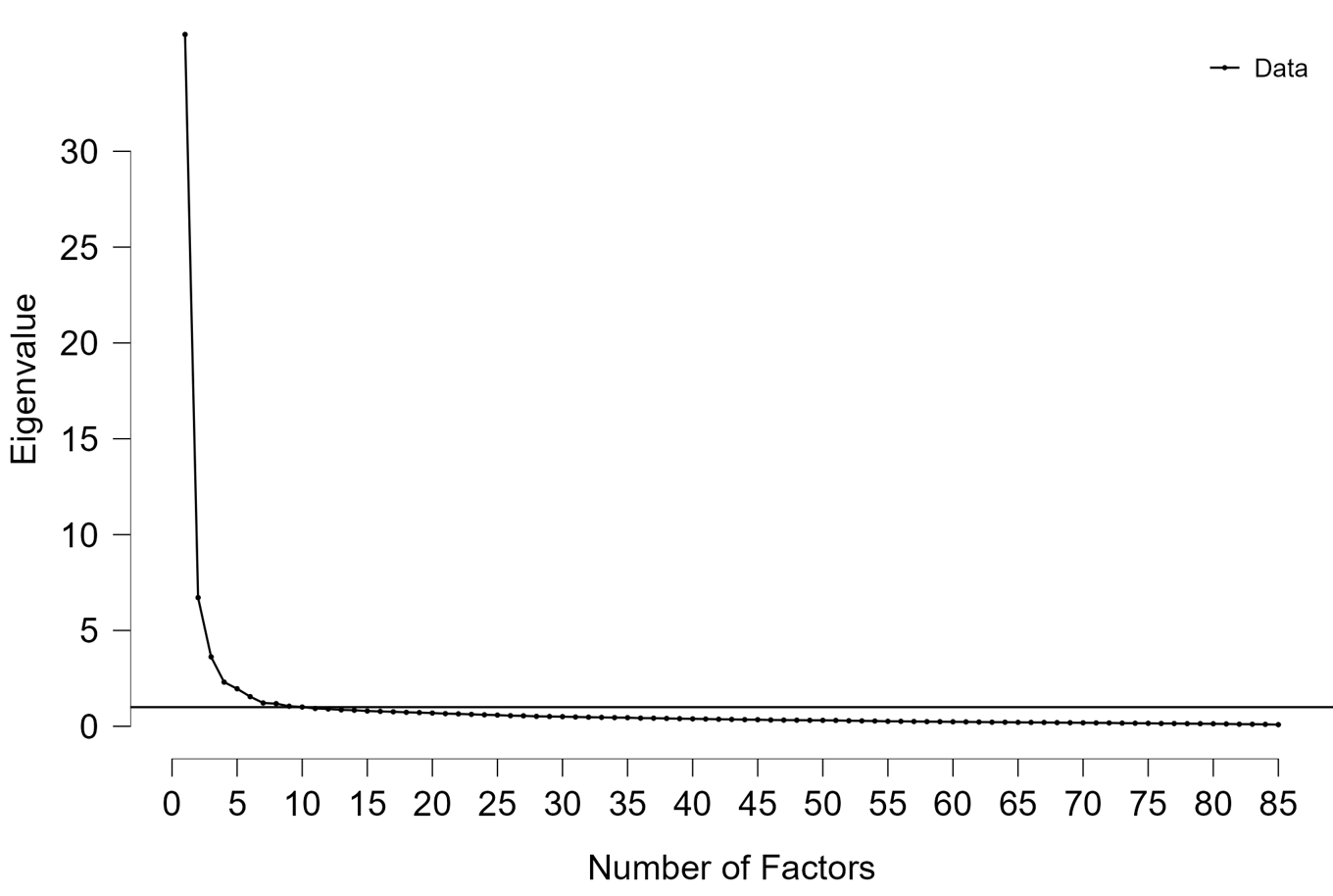}
\caption{Scree plot for EFA model illustrating Eigenvalues versus number of factors.}
\label{fig:2}
\Description{Scree plot for EFA model illustrating Eigenvalues versus number of factors.}
\end{figure*}

\begin{table*}[!t]
\caption{Constructs with Eigenvalue, variance explained, number of items, and Cronbach’s \ensuremath{\alpha}.}
\label{tab:3}
\centering
\begingroup
\small
\setlength{\tabcolsep}{4pt}
\renewcommand{\arraystretch}{1.13}
\setlength{\SEtablewidth}{\dimexpr\linewidth-10\tabcolsep\relax}
\begin{tabular}{@{}>{\raggedright\arraybackslash}p{0.39000\SEtablewidth}>{\centering\arraybackslash}p{0.13000\SEtablewidth}>{\centering\arraybackslash}p{0.11500\SEtablewidth}>{\centering\arraybackslash}p{0.11500\SEtablewidth}>{\centering\arraybackslash}p{0.08500\SEtablewidth}>{\centering\arraybackslash}p{0.16500\SEtablewidth}@{}}
\toprule
\textbf{Latent Constructs} & \textbf{Eigenvalue} & \textbf{Prop. Var.} & \textbf{Cum. Var.} & \textbf{\# Items} & \textbf{Cronbach's \ensuremath{\alpha}} \\
\midrule
1. Requirements Engineering & 35.67 & 14.20\% & 14.20\% & 22 & 0.96 \\
2. Teamwork and Collaboration & 6.32 & 12.40\% & 26.50\% & 15 & 0.95 \\
3. Software Quality Management & 3.20 & 11.50\% & 38.10\% & 16 & 0.95 \\
4. Software Design and Architecture & 1.89 & 9.90\% & 48.00\% & 16 & 0.94 \\
5. Software Agile Methodologies & 1.52 & 9.20\% & 57.20\% & 14 & 0.94 \\
\bottomrule
\end{tabular}
\endgroup
\end{table*}

We used commonly reported fit indices to examine the extent to which the undergraduate computing students’ self-efficacy data fit the final five-factor model: the \textit{\ensuremath{X}}\textsuperscript{2} statistic, the Root Mean Square Error of Approximation (RMSEA), the Tucker-Lewis Index (TLI), the Comparative Fit Index (CFI), and the Standardized Root Mean Square Residual (SRMR). The \textit{\ensuremath{X}}\textsuperscript{2} = 8139.7, \textit{df }= 3155, \textit{p} < .001, RMSEA = 0.055, TLI = 0.832, CFI = 0.852, and SRMR = 0.03. The \textit{\ensuremath{X}}\textsuperscript{2} statistic was statistically significant, which suggest a “lack of fit,” but we remind readers that the \textit{\ensuremath{X}}\textsuperscript{2} statistic is sensitive to sample size, and the present study employs a relatively large sample. As such the \textit{\ensuremath{X}}\textsuperscript{2} statistic should not be solely used to judge the merit of the model. The RMSEA and SRMR suggest a “reasonable fit” based on the criteria of values < .06 and < .08, respectively \cite{Sathyanarayana2024}. Both the RMSEA and SRMR are absolute indices, which assess the fit of the overall model without the need of a baseline null model for reference \cite{Sathyanarayana2024}. However, both the CFI and TLI, which are often labeled as incremental indices, showed that the model did not meet the criteria for “good fit” with > 0.9 and > 0.9, respectively.

To address this conflict, we draw our readers’ attention to our high internal consistency reliability, as measured by Cronbach’s \ensuremath{\alpha}, and to the minimal factor loadings above 0.40 used to assign items to each latent factor, with an average of 0.64 across items. As noted by Groskurth et al. \cite{Groskurth2024}, researchers should use caution in using absolute cutoff scores to justify a model’s fit. As such, we believe all of these pieces of information should be used holistically to make a judgement about the overall fit of the five-factor model on these data. Figure 3 provides the path model for the five-factor solution along with the factor loadings and correlation coefficients among the latent constructs. The visual can be cross-referenced with the pattern matrix in Appendix A and the item-level descriptive statistics found in Appendix B. Readers might recognize that we were able to align our conceptual framing to the factor labels used in the final five-factor solution since the items largely aligned to our theoretical structure.

\begin{figure*}[!t]
\centering
\includegraphics[width=1.0\linewidth]{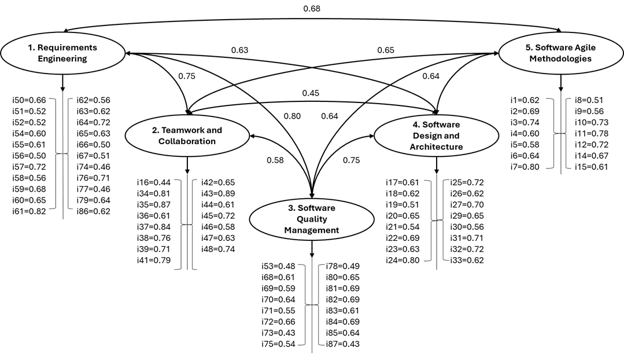}
\caption{Path model for five-factor solution for software engineering self-efficacy constructs.}
\label{fig:3}
\Description{Path model for five-factor solution for software engineering self-efficacy constructs.}
\end{figure*}

\subsection{Descriptive Statistics}

The descriptive statistics from the EFA five-factor model are shown in Table 4. As can be seen, undergraduate computing students rated their highest software engineering self-efficacy in the \textit{Teamwork and Collaboration} sub-construct, with a mean of 4.28 (\textit{SD} = 0.63), followed by \textit{Requirements Engineering} with a mean of 4.03 (\textit{SD} = 0.60). The undergraduate computing student participants rated their self-efficacy in \textit{Software Design and Architecture} as the lowest, with a mean of 3.50 (\textit{SD} = 0.75). It is worth noting that all latent constructs were above the 3.0 central point, indicating that undergraduate computing students have overall higher levels of software engineering self-efficacy. As indicated by the skewness and kurtosis, there were no severe departures from normality, with the values generally within acceptable ranges, with the exception of \textit{Teamwork and Collaboration}, which might be described as a slightly leptokurtic (heavy-tailed) distribution, signifying that the data has a higher peak and fatter tails compared to a normal distribution. The min and max scores show a possible range of four between the two extremes, indicating that at least one participant reported all of the maximum ratings and all of the minimum ratings across the lengthy online survey. Please note that the item-level descriptive statistics are available in Appendix B.

\begin{table*}[!t]
\caption{Descriptive statistics for five-factor EFA model latent constructs.}
\label{tab:4}
\centering
\begingroup
\small
\setlength{\tabcolsep}{4pt}
\renewcommand{\arraystretch}{1.13}
\setlength{\SEtablewidth}{\dimexpr\linewidth-14\tabcolsep\relax}
\begin{tabular}{@{}>{\raggedright\arraybackslash}p{0.34000\SEtablewidth}>{\centering\arraybackslash}p{0.06500\SEtablewidth}>{\centering\arraybackslash}p{0.07000\SEtablewidth}>{\centering\arraybackslash}p{0.08000\SEtablewidth}>{\centering\arraybackslash}p{0.13000\SEtablewidth}>{\centering\arraybackslash}p{0.12000\SEtablewidth}>{\centering\arraybackslash}p{0.09500\SEtablewidth}>{\centering\arraybackslash}p{0.10000\SEtablewidth}@{}}
\toprule
\textbf{Latent Constructs} & \textbf{n} & \textbf{M} & \textbf{SD} & \textbf{Skewness} & \textbf{Kurtosis} & \textbf{Min} & \textbf{Max} \\
\midrule
1. Requirements Engineering & 519 & 4.03 & 0.60 & -0.93 & 3.17 & 1 & 5 \\
2. Teamwork and Collaboration & 522 & 4.28 & 0.63 & -1.61 & 5.27 & 1 & 5 \\
3. Software Quality Management & 519 & 3.76 & 0.71 & -0.67 & 1.43 & 1 & 5 \\
4. Software Design and Architecture & 521 & 3.50 & 0.75 & -0.25 & 0.34 & 1 & 5 \\
5. Software Agile Methodologies & 526 & 3.88 & 0.72 & -0.90 & 1.81 & 1 & 5 \\
\bottomrule
\end{tabular}
\endgroup
\end{table*}

We chose to examine the top five and bottom five individual-item self-efficacy scores to gain further insights into the sample of undergraduate computing students. Table 5 shows the top five highest-rated individual items, which all form the construct of \textit{Teamwork and Collaboration} in the model and manifest different forms of engagement and communication among team members. In contrast, Table 6 shows the five lowest-rated individual items, all of which are contributors to the construct of \textit{Software Design and Architecture}. As noted, the \textit{Software Design and Architecture} construct has the overall lowest score among the five latent factors. Notably, none of the individual item scores were below the 3.0 central point, but it is also worth considering that the standard deviations for the lower-scored items are noticeably larger than those for the highest rated items, indicating higher levels of variability among the undergraduate computing students in our sample.

\begin{table}[!t]
\caption{Top five highest rated individual items on the SESES.}
\label{tab:5}
\centering
\begingroup
\small
\setlength{\tabcolsep}{4pt}
\renewcommand{\arraystretch}{1.13}
\setlength{\SEtablewidth}{\dimexpr\linewidth-6\tabcolsep\relax}
\begin{tabular}{@{}>{\raggedright\arraybackslash}p{0.76000\SEtablewidth}>{\centering\arraybackslash}p{0.08000\SEtablewidth}>{\centering\arraybackslash}p{0.08000\SEtablewidth}>{\centering\arraybackslash}p{0.08000\SEtablewidth}@{}}
\toprule
\textbf{Items} & \textbf{n} & \textbf{M} & \textbf{SD} \\
\midrule
35. I can choose an appropriate channel for formal (e.g., official emails) or informal (e.g., texting) communication & 522 & 4.44 & 0.75 \\
34. I can communicate effectively (both orally and in writing) with different teammates. & 522 & 4.40 & 0.79 \\
43. I can complete my assigned responsibilities on time and communicate any delays to my team & 522 & 4.39 & 0.74 \\
48. I can contribute to building trust and maintaining positive relationships within my team. & 522 & 4.39 & 0.76 \\
37. I can use GitHub for version control and collaborative software development & 522 & 4.38 & 0.75 \\
\bottomrule
\end{tabular}
\endgroup
\end{table}

\begin{table}[!t]
\caption{Bottom five lowest rated individual items on the SESES.}
\label{tab:6}
\centering
\begingroup
\small
\setlength{\tabcolsep}{4pt}
\renewcommand{\arraystretch}{1.13}
\setlength{\SEtablewidth}{\dimexpr\linewidth-6\tabcolsep\relax}
\begin{tabular}{@{}>{\raggedright\arraybackslash}p{0.76000\SEtablewidth}>{\centering\arraybackslash}p{0.08000\SEtablewidth}>{\centering\arraybackslash}p{0.08000\SEtablewidth}>{\centering\arraybackslash}p{0.08000\SEtablewidth}@{}}
\toprule
\textbf{Items} & \textbf{n} & \textbf{M} & \textbf{SD} \\
\midrule
27. I can model N-tier architecture and describe examples of N-tier architecture & 521 & 3.18 & 1.12 \\
31. I can apply the pipe-and-filter architectural style to design systems that process and transform data through sequential stages & 521 & 3.18 & 1.11 \\
26. I can utilize the principles of high cohesion and loose coupling when designing software systems & 521 & 3.28 & 1.13 \\
21. I can understand and explain the differences between Model-View-Controller and Model-View-ViewModel architectures & 521 & 3.30 & 1.21 \\
33. I can design a blackboard architecture for systems that require shared knowledge or iterative problem-solving among components & 521 & 3.34 & 1.10 \\
\bottomrule
\end{tabular}
\endgroup
\end{table}

\subsection{Correlation Analysis}

Table 7 provides the Pearson correlation analyses among the five latent constructs of undergraduate computing students’ software engineering self-efficacy. As shown, all correlation coefficients are positive, statistically significant, and large (\textit{r} > 0.50), except the relationship between \textit{Software Design and Architecture} and \textit{Teamwork and Collaboration}, which is classified as a medium effect size \cite{Gignac2016}. These strong and positive relationships are not surprising, as the EFA model allowed for them (promax) to have relationships, since our working understanding of the multidimensional construct of software engineering self-efficacy aligned to our conceptual framework suggests associations among the latent factors.

The strongest positive relationships identified in the Pearson correlation matrix are between \textit{Requirements Engineering} and \textit{Software Quality Management}, which are two core concepts with several connections in the research literature and curriculum guidelines \cite{Ardis2015}, in that undergraduate computing students who can successfully identify and manage user expectations and requirements, may likely better define the scope of work for a project and more intentionally connect that work to the necessary quality management requirements and processes. The other two strongly and positively related constructs are  between \textit{Teamwork and Collaboration} and \textit{Requirements Engineering}, and \textit{Software Design and Architecture} and \textit{Software Quality Management}. Again, the relationships are meaningful and have likely viable explanations, but we caution our readers to not infer a causal relationship with these findings, as this study does not explain the cause-and-effect between these latent factors.

\begin{table*}[!t]
\caption{Correlation analysis among five-factor model latent constructs.}
\label{tab:7}
\centering
\begingroup
\small
\setlength{\tabcolsep}{4pt}
\renewcommand{\arraystretch}{1.13}
\setlength{\SEtablewidth}{\dimexpr\linewidth-10\tabcolsep\relax}
\begin{tabular}{@{}>{\raggedright\arraybackslash}p{0.50000\SEtablewidth}>{\centering\arraybackslash}p{0.10000\SEtablewidth}>{\centering\arraybackslash}p{0.10000\SEtablewidth}>{\centering\arraybackslash}p{0.10000\SEtablewidth}>{\centering\arraybackslash}p{0.10000\SEtablewidth}>{\centering\arraybackslash}p{0.10000\SEtablewidth}@{}}
\toprule
\textbf{Latent Constructs} & \textbf{1} & \textbf{2} & \textbf{3} & \textbf{4} & \textbf{5} \\
\midrule
1. Requirements Engineering & 1 &  &  &  &  \\
2. Teamwork and Collaboration & 0.75** & 1 &  &  &  \\
3. Software Quality Management & 0.80** & 0.58** & 1 &  &  \\
4. Software Design and Architecture & 0.63** & 0.45** & 0.75** & 1 &  \\
5. Software Agile Methodologies & 0.68** & 0.65** & 0.64** & 0.64** & 1 \\
\bottomrule
\end{tabular}
\par\smallskip
\raggedright * \textit{p} < .01; ** \textit{p} < .001
\endgroup
\end{table*}

\subsection{Individual Differences on Constructs}

As noted, we used a combination of Independent Samples T-tests, ANOVA, and correlation analysis to examine differences in the demographic and background characteristics of undergraduate computing student participants across the software engineering self-efficacy constructs. To execute the models clearly, we collapsed or simplified some of the independent variables into distinct categories to account for the differences. Table 8 provides a summary of the differences by gender (i.e., female versus male), removing the Other (< 2\%) category from the models. As shown, the constructs of \textit{Teamwork and Collaboration}, \textit{Software Design and Architecture}, and \textit{Software Agile Methodologies} show statistically significant differences between males and females. More specifically, females reported higher levels of self-efficacy in \textit{Teamwork and Collaboration} and \textit{Software Agile Methodologies}, while males reported significantly higher levels of self-efficacy in \textit{Software Design and Architecture}.

\begin{table*}[!t]
\caption{Differences based on gender on five-factor model constructs.}
\label{tab:8}
\centering
\begingroup
\small
\setlength{\tabcolsep}{4pt}
\renewcommand{\arraystretch}{1.13}
\setlength{\SEtablewidth}{\dimexpr\linewidth-10\tabcolsep\relax}
\begin{tabular}{@{}>{\raggedright\arraybackslash}p{0.44000\SEtablewidth}>{\centering\arraybackslash}p{0.12000\SEtablewidth}>{\centering\arraybackslash}p{0.12000\SEtablewidth}>{\centering\arraybackslash}p{0.11000\SEtablewidth}>{\centering\arraybackslash}p{0.10000\SEtablewidth}>{\centering\arraybackslash}p{0.11000\SEtablewidth}@{}}
\toprule
\textbf{Latent Constructs} & \textbf{Females} & \textbf{Males} & \textbf{t} & \textbf{df} & \textbf{p} \\
\midrule
1. Requirements Engineering & 4.10 & 4.01 & 1.51 & 508 & 0.132 \\
2. Teamwork and Collaboration & 4.43 & 4.22 & 3.29 & 509 & 0.001 \\
3. Software Quality Management & 3.72 & 3.79 & -1.01 & 508 & 0.314 \\
4. Software Design and Architecture & 3.38 & 3.55 & -2.33 & 508 & 0.020 \\
5. Software Agile Methodologies & 4.03 & 3.83 & 2.68 & 509 & 0.008 \\
\bottomrule
\end{tabular}
\endgroup
\end{table*}

Table 9 illustrates the differences among the student classifications (e.g., freshman, sophomore, junior, and senior), excluding the Non-degree Seeking category (< 0.4\%). We employed Tukey post-hoc follow-up procedures to detect significant differences among the various student classifications that were statistically different. The results show differences in \textit{Software Quality Management}, \textit{Software Design and Architecture}, and \textit{Software Agile Methodologies}. On both \textit{Software Quality Management} and \textit{Software Design and Architecture}, seniors were statistically different than both juniors and sophomores. For \textit{Software Agile Methodologies}, seniors were statistically different than juniors. In each case, seniors had lower self-efficacy scores than their counterparts, which seems counterintuitive.

\begin{table*}[!t]
\caption{Differences based on student classification on five-factor model constructs.}
\label{tab:9}
\centering
\begingroup
\small
\setlength{\tabcolsep}{4pt}
\renewcommand{\arraystretch}{1.13}
\setlength{\SEtablewidth}{\dimexpr\linewidth-12\tabcolsep\relax}
\begin{tabular}{@{}>{\raggedright\arraybackslash}p{0.32000\SEtablewidth}>{\centering\arraybackslash}p{0.09000\SEtablewidth}>{\centering\arraybackslash}p{0.09000\SEtablewidth}>{\centering\arraybackslash}p{0.12500\SEtablewidth}>{\centering\arraybackslash}p{0.14500\SEtablewidth}>{\centering\arraybackslash}p{0.11500\SEtablewidth}>{\centering\arraybackslash}p{0.11500\SEtablewidth}@{}}
\toprule
\textbf{Latent Constructs} & \textbf{F} & \textbf{p} & \textbf{Freshmen} & \textbf{Sophomore} & \textbf{Junior} & \textbf{Senior} \\
\midrule
1. Requirements Engineering & 1.37 & 0.252 & 4.08 & 4.00 & 4.11 & 3.97 \\
2. Teamwork and Collaboration & 0.37 & 0.773 & 4.27 & 4.25 & 4.32 & 4.28 \\
3. Software Quality Management & 6.08 & < .001 & 3.77 & 3.79 & 3.88 & 3.50 \\
4. Software Design and Architecture & 6.88 & < .001 & 3.57 & 3.51 & 3.64 & 3.20 \\
5. Software Agile Methodologies & 5.99 & < .001 & 4.03 & 3.85 & 4.03 & 3.66 \\
\bottomrule
\end{tabular}
\endgroup
\end{table*}

Table 10 presents the differences between students who completed high school computer science courses and those who did not. As can be gleaned, no statistically significant differences were detected across the five latent constructs based on completion of high school computer science courses. Table 11 shows differences in the five latent factors across participants who had or are currently participating in a computing internship programs. Reassuringly, all five factors were statistically significant in favor of those students who have or are currently participating in computing internships. That is, undergraduate computing students' involvement in computing internships appears to significantly and positively influence their self-efficacy in software engineering. This finding is an important discovery that reinforces the critical value of internship experiences for undergraduate computing students.

\begin{table*}[!t]
\caption{Differences based on high school computer science courses on five-factor model constructs.}
\label{tab:10}
\centering
\begingroup
\small
\setlength{\tabcolsep}{4pt}
\renewcommand{\arraystretch}{1.13}
\setlength{\SEtablewidth}{\dimexpr\linewidth-10\tabcolsep\relax}
\begin{tabular}{@{}>{\raggedright\arraybackslash}p{0.44000\SEtablewidth}>{\centering\arraybackslash}p{0.12000\SEtablewidth}>{\centering\arraybackslash}p{0.12000\SEtablewidth}>{\centering\arraybackslash}p{0.12000\SEtablewidth}>{\centering\arraybackslash}p{0.10000\SEtablewidth}>{\centering\arraybackslash}p{0.10000\SEtablewidth}@{}}
\toprule
\textbf{Latent Constructs} & \textbf{t} & \textbf{df} & \textbf{p} & \textbf{No} & \textbf{Yes} \\
\midrule
1. Requirements Engineering & 0.66 & 510 & 0.509 & 4.06 & 4.02 \\
2. Teamwork and Collaboration & 0.24 & 510 & 0.809 & 4.29 & 4.28 \\
3. Software Quality Management & -1.04 & 510 & 0.301 & 3.72 & 3.79 \\
4. Software Design and Architecture & -0.44 & 509 & 0.662 & 3.48 & 3.51 \\
5. Software Agile Methodologies & -0.21 & 510 & 0.834 & 3.88 & 3.89 \\
\bottomrule
\end{tabular}
\endgroup
\end{table*}

\begin{table*}[!t]
\caption{Differences based on participation in computing internship on five-factor model constructs.}
\label{tab:11}
\centering
\begingroup
\small
\setlength{\tabcolsep}{4pt}
\renewcommand{\arraystretch}{1.13}
\setlength{\SEtablewidth}{\dimexpr\linewidth-10\tabcolsep\relax}
\begin{tabular}{@{}>{\raggedright\arraybackslash}p{0.44000\SEtablewidth}>{\centering\arraybackslash}p{0.12000\SEtablewidth}>{\centering\arraybackslash}p{0.12000\SEtablewidth}>{\centering\arraybackslash}p{0.12000\SEtablewidth}>{\centering\arraybackslash}p{0.10000\SEtablewidth}>{\centering\arraybackslash}p{0.10000\SEtablewidth}@{}}
\toprule
\textbf{Latent Constructs} & \textbf{t} & \textbf{df} & \textbf{p} & \textbf{No} & \textbf{Yes} \\
\midrule
1. Requirements Engineering & -3.16 & 510 & 0.002 & 3.98 & 4.16 \\
2. Teamwork and Collaboration & -2.01 & 510 & 0.045 & 4.24 & 4.36 \\
3. Software Quality Management & -4.54 & 510 & < .001 & 3.67 & 3.97 \\
4. Software Design and Architecture & -3.89 & 509 & < .001 & 3.41 & 3.68 \\
5. Software Agile Methodologies & -3.45 & 510 & < .001 & 3.81 & 4.04 \\
\bottomrule
\end{tabular}
\endgroup
\end{table*}

We converted the number of completed college computing courses into an ordinal scale (i.e., one through six) and correlated the scores using Spearman’s rho across the five latent constructs. We employed Spearman’s rho since the number of completed college computing courses is measured on an ordinal scale. Table 12 shows the correlations between the number of previously completed college computing courses by the student participants and the five latent factors. As shown, two variables were statistically significant: an undergraduate student’s prior number of college computing courses completed and \textit{Teamwork and Collaboration} and \textit{Requirements Engineering}, both of which show a small, positive relationship. While Software Quality Management was approaching significance, it did not meet the \textit{p} < .05 criterion. It appears that the number of prior college computing courses completed does not have a discernible relationship with the software engineering self-efficacy latent constructs. This finding is counter-intuitive as one would expect undergraduate computing students with more expertise to report higher levels of software engineering self-efficacy.

\begin{table}[!t]
\caption{Spearman’s rho correlations between five latent factors and the number of previous college computing courses completed.}
\label{tab:12}
\centering
\begingroup
\small
\setlength{\tabcolsep}{4pt}
\renewcommand{\arraystretch}{1.13}
\setlength{\SEtablewidth}{\dimexpr\linewidth-4\tabcolsep\relax}
\begin{tabular}{@{}>{\raggedright\arraybackslash}p{0.72000\SEtablewidth}>{\centering\arraybackslash}p{0.14000\SEtablewidth}>{\centering\arraybackslash}p{0.14000\SEtablewidth}@{}}
\toprule
\textbf{Latent Constructs} & \textbf{r}\textbf{\textsubscript{s}} & \textbf{p} \\
\midrule
1. Requirements Engineering & 0.12 & 0.01 \\
2. Teamwork and Collaboration & 0.11 & 0.01 \\
3. Software Quality Management & 0.09 & 0.05 \\
4. Software Design and Architecture & -0.01 & 0.89 \\
5. Software Agile Methodologies & -0.01 & 0.83 \\
\bottomrule
\end{tabular}
\endgroup
\end{table}

Our final Table 13 shows the race of the undergraduate computing student participants by the five software engineering self-efficacy latent constructs. As shown, two latent factors were statistically significant among the differences in the race categories in the models: \textit{Software Quality Management} and \textit{Software Agile Methodologies}. In the case of \textit{Software} \textit{Quality Management}, Asians differed statistically from Mixed-race undergraduate computing students, with Asians reporting higher self-efficacy scores than their Mixed-race counterparts. In the construct of \textit{Software Agile Methodologies}, White/Caucasians differed statistically from Asians, again with Asians reporting higher self-efficacy scores compared to their White/Caucasians peers. No other statistically significant differences were detected in these models across the other race categories. Notable, as previously shown, Asians represented approximately 46\% of the undergraduate computing students who participated in the study.

\begin{table*}[!t]
\caption{Differences based on race on five-factor model constructs.}
\label{tab:13}
\centering
\begingroup
\small
\setlength{\tabcolsep}{4pt}
\renewcommand{\arraystretch}{1.13}
\setlength{\SEtablewidth}{\dimexpr\linewidth-14\tabcolsep\relax}
\begin{tabular}{@{}>{\raggedright\arraybackslash}p{0.32000\SEtablewidth}>{\centering\arraybackslash}p{0.07500\SEtablewidth}>{\centering\arraybackslash}p{0.08500\SEtablewidth}>{\centering\arraybackslash}p{0.12000\SEtablewidth}>{\centering\arraybackslash}p{0.10000\SEtablewidth}>{\centering\arraybackslash}p{0.10000\SEtablewidth}>{\centering\arraybackslash}p{0.10000\SEtablewidth}>{\centering\arraybackslash}p{0.10000\SEtablewidth}@{}}
\toprule
\textbf{Latent Constructs} & \textbf{F} & \textbf{p} & \textbf{White/Cauc} & \textbf{Asian} & \textbf{Hisp/Lat} & \textbf{Mixed} & \textbf{Black/AA} \\
\midrule
1. Requirements Engineering & 0.49 & 0.746 & 4.06 & 4.05 & 4.06 & 3.94 & 3.99 \\
2. Teamwork and Collaboration & 0.94 & 0.44 & 4.29 & 4.28 & 4.43 & 4.24 & 4.12 \\
3. Software Quality Management & 3.18 & 0.013 & 3.72 & 3.87 & 3.71 & 3.58 & 3.52 \\
4. Software Design and Architecture & 1.63 & 0.166 & 3.45 & 3.59 & 3.37 & 3.42 & 3.35 \\
5. Software Agile Methodologies & 2.80 & 0.025 & 3.76 & 4.00 & 3.88 & 3.80 & 3.76 \\
\bottomrule
\end{tabular}
\endgroup
\end{table*}

\section{Discussion}

\subsection{Limitations and Delimitations}

Interpretation of our research findings should be framed within the limitations and delimitations of the present study. First, we collected cross-sectional data from undergraduate computing students at three institutions of higher education in the southeastern U.S. While we recruited enough participants to meet the scientific objectives of this study, readers should be cautious about generalizing these findings to other regions of the U.S. or outside the U.S. Second, participants were recruited from SE courses as well as subsequent computing courses, meaning that some students may have completed the focal SE course one, two, or more semesters prior to participating in this study, which could have influenced the accuracy of their input and perceptions. Third, although we generated a research-inspired conceptual framework to operationalize software engineering self-efficacy among undergraduate computing students, this framework was intentionally aligned to the software engineering courses at two of the academic institutions. Thus, our theory may not be comprehensive enough to account for all contemporary approaches to software engineering. Fourth, the concept of self-efficacy used in this study accounted only for the mastery experience dimension of self-efficacy, which does not encompass the other dimensions (e.g., vicarious experiences, social persuasion, and emotional/physiological states) outlined by Bandura and other scholars \cite{Bandura1977}. Although self-efficacy measures often focus narrowly on mastery experiences \cite{Maurer1998} using “I can” statements, we acknowledge that this approach captures only a single dimension of the complex construct. Finally, the SESES is a self-report measure of software engineering self-efficacy and, as such, is subject to common biases, such as social desirability bias, recall bias, and poor self-awareness bias \cite{Podsakoff2024}. In light of these limitations and delimitations, our findings provide some useful insights into an undergraduate computing student’s software engineering self-efficacy and the preliminary psychometric properties of the SESES.

\subsection{Software Engineering Self-Efficacy}

So what can we conclude from our research study about the nature of software engineering self-efficacy among undergraduate computing students? First, we have provided strong evidence that our conceptual framework, inspired by the notion of mastery experience from self-efficacy \cite{Bandura1977} and by contemporary literature and the computing curriculum guidance on software engineering, may accurately portray undergraduate computing students’ software engineering self-efficacy. Our findings provide strong preliminary evidence that software engineering self-efficacy among undergraduate computing students is likely a multidimensional construct with strong inter-relationships among the constructs that manifest this complex idea. While we fully acknowledge that our conception of software engineering self-efficacy may not be inclusive of all related concepts (e.g., software engineering cost management) and approaches to the research and practice of software engineering (e.g., little coverage of traditional software development lifecycles inspired by the waterfall method) \cite{Kitchenham2005}, we feel this conceptual framework and the associated SESES is a strong starting place for future research on this moving target in the computing domain.

We examined the wide range of topics, knowledge, and skills expected of undergraduate computing students as conceived in our conceptual framework and the design of the SESES, and we cannot help but wonder whether we are expecting too much from our “typical” undergraduate computing students. Many of the concepts covered within undergraduate software engineering courses are difficult for undergraduate students to grasp without extensive experience working with diverse teams, and more often than not, working on complex and large existing code bases written in languages and deployed on platforms often no longer covered in modern computing curricula (e.g., LAMP stack, COBOL, etc) \cite{MainframesND}. Software engineering concepts may only be appreciated as our undergraduates enter the profession and gain several years of firsthand experience with these ideas. As computing educators and researchers, it is incumbent upon us to openly discuss when our curriculum may not realistically create learning experiences that can fully achieve these multidimensional learning outcomes, while also ensuring students are capable of meeting the growing expectations (e.g., soft skills) of employers. With the advent of GenAI and other forms like agentic AI, computing educators and researchers are at a pivotal moment in our several decades of existence.

We found it quite telling that the highest-rated software engineering self-efficacy construct was that of \textit{Teamwork and Collaboration}. While it is common practice for computing educators to create meaningful group projects to nurture the development of these skills, employers are often the first to note that undergraduates lack them \cite{Garousi2019}. This mismatch between employers' and students’ self-efficacy may be evidence of a poor self-awareness bias \cite{DunningKruger2011}, in which computing undergraduates believe they are capable of working effectively in diverse teams but, in practice, often fall short of the benchmarks and expectations set by modern employers. We also note that \textit{Requirements Engineering} was the second-highest-rated self-efficacy construct, yet we question whether undergraduate students are truly afforded deep learning experiences that demand careful engagement with real-world stakeholders to elicit, document, verify, and deploy their technological business needs. For instance, computing curricula often involve undergraduate students creating and deploying their own ideas as computing solutions, using tools of their choice or shaped by the educator or institution, and inventing requirements, perhaps in negotiation with an educator or in collaboration with their peers \cite{Iacob2019}. This is simply not how \textit{Requirements Engineering} is executed in business and other organizational settings, and again is evidence of a mismatch between undergraduate students' software engineering self-efficacy and their actual mastery experiences.

The two lowest-rated software engineering self-efficacy constructs were \textit{Software Quality Management} and \textit{Software Design and Architecture}, which are strongly related, as indicated by our findings. While an undergraduate computing student can study and even potentially deploy some common design patterns or system architectures in their academic work \cite{Lartigue2018}, the connection between these concepts and the ability to evaluate trade-offs among concepts like reusability, scalability, usability, and performance may be outside the reach of what is realistically possible in a traditional 16-week academic semester. Computing educators and researchers alike should seek to balance curriculum expectations with students' ability to master the concepts, both during their academic program and in real-world experiences such as internships or work-study, so undergraduates have ample opportunities to apply these ideas in practice and observe firsthand the trade-offs and decisions software engineers face in their workplace. Many ideas presented to undergraduates in a software engineering course are not fully appreciated or understood until later in their professional careers \cite{Karunasekera2007}; thus, we must identify which concepts are best covered before, during, and after the software engineering course in a computing curriculum to ensure the concepts come to fruition at the time when the concepts are needed in their practice.

Software engineering self-efficacy is moderated by several characteristics among the undergraduate computing students in our sample. For instance, females had overall higher scores in the constructs of \textit{Teamwork and Collaboration} and \textit{Software Agile Methodologies}, which can both be characterized as people-centered skills, while males had higher scores in the construct of \textit{Software Design and Architecture}, a far more technical skill set. While gender stereotypes are problematic to the computer science discipline, empirical research continues to show notable differences between male and female counterparts \cite{Yates2022} in the computing discipline. Another interesting difference was that undergraduate computing students who had participated in computing internships consistently had higher software engineering self-efficacy scores across the five constructs, which provides further empirical justification for students to pursue internship opportunities during their studies. Computing internships not only enhance the students' ability to secure a full-time position upon graduation \cite{Wolf2023} but also appear to inherently boost their level of software engineering self-efficacy, which can translate to persistence and longer-term goals and commitments to the computing profession \cite{Ojha2024}. Other differences were detected in our models, warranting further investigation within the computing education research community.

To our knowledge, the SESES is the first attempt in the computing education research literature to operationalize and measure an undergraduate computing student’s software engineering self-efficacy as a target multidimensional construct. The conceptual framework used to create the initial item pool was grounded in contemporary software engineering literature and curriculum guidelines, and it was carefully aligned with the learning experience in two software engineering courses from different universities. Using guidance from Classical Test theory, we used a systematic process to collect validity evidence (e.g., content validity) to support the use of the  SESES in computing education research, evaluation, and practice. The SESES yielded a five-factor model aligned with our conceptual framework, with highly internally consistent latent factors, simple structures in the pattern matrix, and meaningful and explainable factor loadings. Our absolute fit indices indicated a “good fit,” but our incremental fit indices did not, a well-documented phenomenon in educational measurement \cite{Groskurth2024}. At this stage, the SESES provides preliminary validity and reliability evidence to support its use for low-stakes purposes, but future research is needed to extend the conceptual model to other undergraduate computing populations and to test the theoretical model using a Confirmatory Factor Analysis (CFA).

\subsection{Implications for Practice}

Computing educators are in a precarious situation due to recent advances in GenAI models and their rapid adoption in the software development life cycle, often referred to as “vibe coding” \cite{Kusper2025}. Our entire curriculum sequence is firmly rooted in the foundations of computer programming, using an incremental approach across multiple layers of coursework (e.g., CS1, CS2, and data structures and algorithms), in which undergraduate computing students' primary learning activity is the “programming assignment” used to master the learning outcomes. With GenAI models and tools capable of solving any authentic programming assignment we can dream up that is developmentally appropriate for introductory and intermediate undergraduate computing students, we are now challenged to assess our students when they can use a GenAI model in a matter of seconds with clever prompting to solve our masterfully designed programming assignments that carefully trace the student learning outcomes in our courses. We now face a reckoning and disruption on rethinking our teaching and assessment methods - felt throughout higher education - in jeopardizing our foundational identity and reputation as a discipline, one long-coveted for career-readiness, academic rigor and relevance, and higher-than-average starting salaries for our graduates \cite{Dickler2026}. For long, the software engineering course has been a pinnacle of the undergraduate computing students’ academic journey, in which computing educators create collaborative, authentic, and creative learning experiences that allow undergraduates to use the vast knowledge and skills gained from their earlier coursework to solve real-world challenges. Yet, now computing educators face the insurmountable challenge of choosing when and how to integrate GenAI models into the curriculum in an ethical, transparent, and responsible way without jeopardizing the critical concepts embedded in our learning outcomes.

A tool like the SESES can be used by computing educators in many useful ways to ensure the learning outcomes are achieved and enhanced from semester to semester. The SESES can be used as a pre-learning activity to gauge undergraduate computing students’ current self-efficacy in software engineering. Doing so could serve the simultaneous purpose of measuring students’ current levels while also priming them for the expectations in the software engineering course. The tool could also be used as a post-assessment to measure changes in the five model factors after completing the activities in a software engineering course. The information gathered from the SESES can be used by computing educators to tailor lessons to meet the needs of their current students and to inform future curriculum and instructional modifications to better serve their future undergraduate computing students. The SESES could also be applied to the evaluation of various educational interventions and programs focused on software engineering outcomes, to make data-based decisions about the educational utility and efficacy of different learning resources used in the curriculum.

\subsection{Implications for Future Research}

While the present study provides a preliminary framework for the use of SESES in software engineering education research, the next step is to validate the theoretical model across different conditions, populations, and learning sequences. That is, following the guidance of Classical Test theory \cite{Crocker1986}, we need to gather additional evidence from a different, robust sample of the target population and conduct additional analyses, such as a CFA and even differential item functioning (DIF), to test for bias at the individual item level. The SESES could be used in several applications in educational computing research, including descriptive and comparative studies of targeted educational programs focused on software engineering learning outcomes, as well as in larger computing education initiatives in both formal (e.g., traditional courses) and informal learning environments (e.g., hackathons or online microcredential programs). The measure could be employed in longitudinal studies to track student progress in key software engineering areas over time. This type of study with the SESES would be particularly useful in generating test-retest reliability evidence since the SESES would be administered on several occasions with the same participants, which generates reliability evidence beyond the mere internal consistency coefficients. Likewise, the SESES could be repurposed for applications in the business and industry settings for talent development initiatives such as planning targeted professional development opportunities or self-assessing an employee’s software engineering self-efficacy to better align their beliefs and competencies. With appropriate modifications and additional data collection, the SESES could be repurposed for other use cases so long as data is collected with a new target population to further validate the responsible and ethical use of the scale.

There is also the need to provide stronger forms of convergent and discriminant validity evidence using complimentary and unrelated measurements, such as using the established multi-trait, multi-method approach \cite{Campbell1959}, in which similar or dissimilar constructs can be measured using different methods (e.g., observation rubrics or artifact analysis) with aim of providing strong validity arguments to support the use of a measurement tool. The software engineering self-efficacy scores can and should be correlated with undergraduate computing students’ academic performance scores (i.e., overall or by type of assessment outcome, such as group project performance, quizzes, or examinations) upon the conclusion of a software engineering course to potentially illustrate the more robust predictive power of this domain-specific measure. This form of evidence would enhance the use cases for the SESES across formal educational settings serving computing undergraduate students by providing the level of evidence necessary for wide-spread adoption among computing educators and researchers alike. As no established measurement systems are available for undergraduate computing students' software engineering self-efficacy, the SESES serves as a starting point for future contributions to the domain-specific measure of software engineering self-efficacy. While there are clearly noted limitations and delimitations to this study, the findings provide a foundation for future research on contributing to the creation of computing self-efficacy scales capable of measuring the domain-specific context of our discipline.

Returning to our three guiding research questions, the present study offers preliminary answers that warrant further investigation. Regarding the underlying latent constructs of undergraduate computing student’s software engineering self-efficacy (RQ1), our findings support a multidimensional conceptualization comprising five interrelated constructs (Teamwork and Collaboration, Requirements Engineering, Software Agile Methodologies, Software Quality Management, and Software Design and Architecture) aligned with our research-inspired conceptual framework. Regarding the validity and reliability evidence supporting the SESES (RQ2), our analyses yielded a five-factor model with highly internally consistent latent factors, a simple structure in the pattern matrix, and meaningful factor loadings, alongside absolute fit indices indicating a good fit; these results provide preliminary validity and reliability evidence to support the use of the SESES for low-stakes purposes, while further evidence, particularly from Confirmatory Factor Analysis, remains needed. Regarding differences across undergraduate computing students' demographics and backgrounds (RQ3), software engineering self-efficacy was moderated by several characteristics, with notable differences observed by gender across specific constructs and consistently higher scores among students who had participated in computing internships. Taken together, these findings position the SESES as a preliminary but promising starting point for future research on software engineering self-efficacy in computing education.

\section*{Acknowledgments}
The authors thank all individuals who voluntarily participated in this study. Their input was essential for the outcomes reported in this manuscript.

This material is based upon work supported by the National Science Foundation under Grant No. 2434428. Any opinions, findings, and conclusions or recommendations expressed in this material are those of the author(s) and do not necessarily reflect the views of the National Science Foundation.

\bibliographystyle{IEEEtran}

\nocite{*}

\bibliography{references_ToE}

\onecolumn
\appendices

\section{}

\begingroup
\small
\setlength{\tabcolsep}{4pt}
\renewcommand{\arraystretch}{1.05}
\setlength{\SEtablewidth}{\dimexpr\linewidth-10\tabcolsep\relax}
\begin{longtable}{@{}>{\raggedright\arraybackslash}p{0.60000\SEtablewidth}>{\centering\arraybackslash}p{0.08000\SEtablewidth}>{\centering\arraybackslash}p{0.08000\SEtablewidth}>{\centering\arraybackslash}p{0.08000\SEtablewidth}>{\centering\arraybackslash}p{0.08000\SEtablewidth}>{\centering\arraybackslash}p{0.08000\SEtablewidth}@{}}
\toprule
\textbf{Latent Constructs/Items} & \textbf{Factor 1} & \textbf{Factor 2} & \textbf{Factor 3} & \textbf{Factor 4} & \textbf{Factor 5} \\
\midrule
\endfirsthead
\toprule
\textbf{Latent Constructs/Items} & \textbf{Factor 1} & \textbf{Factor 2} & \textbf{Factor 3} & \textbf{Factor 4} & \textbf{Factor 5} \\
\midrule
\endhead
\midrule
\endfoot
\bottomrule
\endlastfoot
\addlinespace[4pt]
\multicolumn{6}{@{}l@{}}{\textbf{1. }\textbf{\textit{Requirements Engineering}}} \\*
61. I can prioritize requirements based on user and project goals. & 0.82 &  &  &  &  \\
64. I can check that implemented features meet stakeholder expectations. & 0.721 &  &  &  &  \\
57. I can identify key objects and relationships from the requirement documentation. & 0.718 &  &  &  &  \\
76. I can uphold ethical and professional standards when working on software projects & 0.713 &  &  &  &  \\
59. I can gather user and stakeholder needs through interviews or workshops. & 0.678 &  &  &  &  \\
50. I can break down requirements into modules and software development tasks. & 0.659 &  &  &  &  \\
60. I can write use cases and user stories that reflect real user needs. & 0.651 &  &  &  &  \\
79. I can apply lessons learned from previous projects to enhance quality in future work. & 0.641 &  &  &  &  \\
65. I can manage requirement changes using documentation or version control tools. & 0.629 &  &  &  &  \\
86. I can apply ethical principles when making software design decisions. & 0.62 &  &  &  &  \\
63. I can connect requirements to code, design, and testing artifacts. & 0.619 &  &  &  &  \\
55. I can find missing or conflicting parts in requirement specifications. & 0.608 &  &  &  &  \\
54. I can adapt existing code to accommodate changing requirements. & 0.602 &  &  &  &  \\
58. I can translate complex requirements into clear, usable documentation. & 0.558 &  &  &  &  \\
62. I can update requirement documents when project conditions change. & 0.557 &  &  &  &  \\
51. I can design a system architecture that satisfies the given requirements. & 0.516 &  &  &  &  \\
52. I can write clean, well-documented, and maintainable code that complies with team standards. & 0.515 &  &  &  &  \\
67. I can keep requirements traceable throughout the project lifecycle. & 0.511 &  &  &  &  \\
56. I can model requirements using appropriate diagrams (e.g., entity-relationship) that clearly show system characteristics and behaviors. & 0.5 &  &  &  &  \\
66. I can negotiate priorities and trade-offs with relevant stakeholders. & 0.496 &  &  &  &  \\
74. I can conduct and contribute effectively to peer code reviews with my teammates & 0.464 &  &  &  &  \\
77. I can identify and prioritize risks that impact software quality across different contexts & 0.459 &  &  &  &  \\
\addlinespace[4pt]
\multicolumn{6}{@{}l@{}}{\textbf{2. }\textbf{\textit{Teamwork and Collaboration}}} \\*
43. I can complete my assigned responsibilities on time and communicate any delays to my team &  & 0.892 &  &  &  \\
35. I can choose an appropriate channel for formal (e.g., official emails) or informal (e.g., texting) communication &  & 0.872 &  &  &  \\
37. I can use GitHub for version control and collaborative software development &  & 0.838 &  &  &  \\
34. I can communicate effectively (both orally and in writing) with different teammates. &  & 0.809 &  &  &  \\
41. I can take the initiative to organize meetings, progress checks, and help my teammates meet deadlines &  & 0.792 &  &  &  \\
38. I can manage branches, commits, and pull requests on GitHub &  & 0.761 &  &  &  \\
48. I can contribute to building trust and maintaining positive relationships within my team. &  & 0.74 &  &  &  \\
45. I can collaborate effectively with others to resolve merge or integration conflicts &  & 0.724 &  &  &  \\
39. I can use issues, comments, and notifications on GitHub for coordination &  & 0.712 &  &  &  \\
42. I can mediate disagreements among team members &  & 0.647 &  &  &  \\
47. I can support my teammates when they face technical (e.g., runtime errors) or organizational (e.g., conflicts with other team members) challenges. &  & 0.63 &  &  &  \\
36. I can collaboratively plan user stories to organize project responsibilities and coordinate coding tasks. &  & 0.613 &  &  &  \\
44. I can review my teammates' code and provide constructive, respectful suggestions to improve their work &  & 0.612 &  &  &  \\
46. I can present and pitch ideas for professional software development environments &  & 0.577 &  &  &  \\
16. I can adapt to change when new requirements are identified &  & 0.441 &  &  &  \\
\addlinespace[4pt]
\multicolumn{6}{@{}l@{}}{\textbf{3. }\textbf{\textit{Software Quality Management}}} \\*
82. I can evaluate software performance and stability across diverse usage environments. &  &  & 0.689 &  &  \\
84. I can design software components that protect users' data privacy and security. &  &  & 0.688 &  &  \\
81. I can implement techniques to improve reliability and fault tolerance under different conditions. &  &  & 0.686 &  &  \\
72. I can automate and execute tests to verify software behavior across multiple software environments &  &  & 0.663 &  &  \\
80. I can detect and mitigate potential security vulnerabilities across different platforms and technologies. &  &  & 0.652 &  &  \\
85. I can select and manage appropriate cloud-based platforms (SaaS, PaaS, IaaS) for software deployment. &  &  & 0.637 &  &  \\
70. I can utilize quality analysis tools to assess and improve code health across various software systems &  &  & 0.637 &  &  \\
68. I can recognize and address code smells or design flaws in different programming environments &  &  & 0.614 &  &  \\
83. I can identify and refactor code to improve its quality and maintainability. &  &  & 0.607 &  &  \\
69. I can refactor existing code to enhance its structure and maintainability across various software systems &  &  & 0.585 &  &  \\
71. I can create effective test cases from specifications or existing software across different contexts &  &  & 0.547 &  &  \\
75. I can use software engineering metrics to evaluate and monitor quality across multiple projects &  &  & 0.535 &  &  \\
78. I can use version control, Continuous Integration/Continuous Delivery, or other tools to maintain continuous quality over time. &  &  & 0.494 &  &  \\
53. I can prepare and execute deployment procedures (i.e., releasing software components into production). &  &  & 0.477 &  &  \\
87. I can utilize Development and Operations (DevOps) practices, such as continuous integration and automated testing, to enhance software quality. &  &  & 0.431 &  &  \\
73. I can apply standard coding, documentation, and process guidelines consistently in diverse settings &  &  & 0.429 &  &  \\
\addlinespace[4pt]
\multicolumn{6}{@{}l@{}}{\textbf{4. }\textbf{\textit{Software Design and Architecture}}} \\*
24. I can design event-driven systems and describe the varying subtypes of an event-driven architecture &  &  &  & 0.804 &  \\
32. I can justify architectural design decisions by documenting the rationale and the attributes they address &  &  &  & 0.724 &  \\
25. I can model peer-to-peer architecture and describe examples of peer-to-peer architecture &  &  &  & 0.724 &  \\
31. I can apply the pipe-and-filter architectural style to design systems that process and transform data through sequential stages &  &  &  & 0.712 &  \\
27. I can model N-tier architecture and describe examples of N-tier architecture &  &  &  & 0.702 &  \\
22. I can understand the uses of layered software architecture systems (e.g., Presentation, Business, Data layers) &  &  &  & 0.69 &  \\
20. I can describe and implement systems using common architectures, such as client-server, peer-to-peer, N-tier, and pipe-and-filter. &  &  &  & 0.65 &  \\
29. I can analyze design trade-offs (e.g., performance vs. scalability) when selecting a software architecture &  &  &  & 0.648 &  \\
23. I can create the appropriate diagram (e.g., entity-relationship) to represent a system I am developing &  &  &  & 0.627 &  \\
33. I can design a blackboard architecture for systems that require shared knowledge or iterative problem-solving among components &  &  &  & 0.62 &  \\
26. I can utilize the principles of high cohesion and loose coupling when designing software systems &  &  &  & 0.619 &  \\
18. I can determine which software architecture will best fit a project &  &  &  & 0.615 &  \\
17. I can differentiate between software archetypes based on performance, scalability, cost, and other factors. &  &  &  & 0.609 &  \\
30. I can decompose a complex software system into components and modules that promote maintainability and reuse &  &  &  & 0.56 &  \\
21. I can understand and explain the differences between Model-View-Controller and Model-View-ViewModel architectures &  &  &  & 0.54 &  \\
19. I can create appropriate diagrams (e.g., flowcharts) to properly model systems that I build &  &  &  & 0.509 &  \\
\addlinespace[4pt]
\multicolumn{6}{@{}l@{}}{\textbf{5. }\textbf{\textit{Software Agile Methodologies}}} \\*
7. I can refactor existing code to enhance its structure and maintainability across various software systems &  &  &  &  & 0.803 \\
11. I understand how Agile philosophy can be utilized to form various methodologies &  &  &  &  & 0.783 \\
3. I can analyze design trade-offs (e.g., performance vs. scalability) when selecting a software architecture &  &  &  &  & 0.738 \\
10. I can put the principles of agile development to use in my work &  &  &  &  & 0.734 \\
12. I can understand the difference between agile and waterfall methodologies &  &  &  &  & 0.723 \\
2. I can create appropriate diagrams (e.g., flowcharts) to properly model systems that I build &  &  &  &  & 0.688 \\
14. I can identify the utility of the various artifacts of agile &  &  &  &  & 0.666 \\
6. I can gather user and stakeholder needs through interviews or workshops. &  &  &  &  & 0.642 \\
1. I can put the principles of agile development to use in my work &  &  &  &  & 0.622 \\
15. I can think critically about tradeoffs (e.g., flexibility vs. quick delivery) in a Scrum environment &  &  &  &  & 0.614 \\
4. I can use issues, comments, and notifications on GitHub for coordination &  &  &  &  & 0.599 \\
5. I can plan and estimate the effort and time required for software development tasks. &  &  &  &  & 0.582 \\
9. I can utilize Development and Operations (DevOps) practices, such as continuous integration and automated testing, to enhance software quality. &  &  &  &  & 0.558 \\
8. I can apply lessons learned from previous projects to enhance quality in future work. &  &  &  &  & 0.506 \\
\addlinespace[4pt]
\multicolumn{6}{@{}l@{}}{\textbf{\textit{Dropped from Model Due to Low Factor Loadings}}} \\*
13. I can deliver measurable value for customers through my work &  &  &  &  &  \\
49. I can plan and estimate the effort and time required for software development tasks. &  &  &  &  &  \\
\addlinespace[4pt]
\multicolumn{6}{@{}l@{}}{\textbf{\textit{Removed from Model Due to Cross-Loadings}}} \\*
28. I can describe and utilize the core ideas of object-oriented programming when designing software systems. &  &  &  &  &  \\
40. I can troubleshoot Git-related problems (e.g., merge conflicts) in team settings. &  &  &  &  &  \\
\end{longtable}
\endgroup

\clearpage

\section{}

\begingroup
\small
\setlength{\tabcolsep}{4pt}
\renewcommand{\arraystretch}{1.05}
\setlength{\SEtablewidth}{\dimexpr\linewidth-6\tabcolsep\relax}
\begin{longtable}{@{}>{\raggedright\arraybackslash}p{0.76000\SEtablewidth}>{\centering\arraybackslash}p{0.08000\SEtablewidth}>{\centering\arraybackslash}p{0.08000\SEtablewidth}>{\centering\arraybackslash}p{0.08000\SEtablewidth}@{}}
\toprule
\textbf{Items} & \textbf{\textit{n}} & \textbf{\textit{M}} & \textbf{\textit{SD}} \\
\midrule
\endfirsthead
\toprule
\textbf{Items} & \textbf{\textit{n}} & \textbf{\textit{M}} & \textbf{\textit{SD}} \\
\midrule
\endhead
\midrule
\endfoot
\bottomrule
\endlastfoot
1. I can put the principles of agile development to use in my work & 527 & 3.89 & 0.88 \\
2. I can work well within a Scrum environment & 527 & 3.97 & 0.88 \\
3. I can understand the responsibilities of a Scrum master & 526 & 3.97 & 0.92 \\
4. I can organize a sprint for my team & 526 & 3.99 & 0.91 \\
5. I can fulfill my role in a daily standup & 526 & 4.16 & 0.84 \\
6. I can maintain a healthy backlog (e.g., 1-2 sprints) on a project & 526 & 3.96 & 0.84 \\
7. I can identify and differentiate between the steps of the Agile development Lifecycle & 526 & 3.69 & 1.03 \\
8. I can utilize a board to keep track of the progress of my team's ongoing work & 526 & 4.11 & 0.86 \\
9. I can fulfill my role as a developer in a sprint planning meeting & 526 & 4.07 & 0.88 \\
10. I can differentiate between Scrum and other Agile methods & 526 & 3.53 & 1.11 \\
11. I understand how Agile philosophy can be utilized to form various methodologies & 526 & 3.74 & 1.00 \\
12. I can understand the difference between agile and waterfall methodologies & 526 & 3.60 & 1.21 \\
13. I can deliver measurable value for customers through my work & 526 & 4.07 & 0.82 \\
14. I can identify the utility of the various artifacts of agile & 526 & 3.68 & 1.02 \\
15. I can think critically about tradeoffs (e.g., flexibility vs. quick delivery) in a Scrum environment & 526 & 3.96 & 0.89 \\
16. I can adapt to change when new requirements are identified & 526 & 4.13 & 0.80 \\
17. I can differentiate between software archetypes based on performance, scalability, cost, and other factors. & 522 & 3.64 & 0.96 \\
18. I can determine which software architecture will best fit a project & 521 & 3.52 & 0.91 \\
19. I can create appropriate diagrams (e.g., flowcharts) to properly model systems that I build & 521 & 3.87 & 0.87 \\
20. I can describe and implement systems using common architectures, such as client-server, peer-to-peer, N-tier, and pipe-and-filter. & 521 & 3.39 & 1.06 \\
21. I can understand and explain the differences between Model-View-Controller and Model-View-ViewModel architectures & 521 & 3.30 & 1.21 \\
22. I can understand the uses of layered software architecture systems (e.g., Presentation, Business, Data layers) & 521 & 3.57 & 0.99 \\
23. I can create the appropriate diagram (e.g., entity-relationship) to represent a system I am developing & 521 & 3.67 & 0.94 \\
24. I can design event-driven systems and describe the varying subtypes of an event-driven architecture & 521 & 3.51 & 1.03 \\
25. I can model peer-to-peer architecture and describe examples of peer-to-peer architecture & 521 & 3.45 & 1.04 \\
26. I can utilize the principles of high cohesion and loose coupling when designing software systems & 521 & 3.28 & 1.13 \\
27. I can model N-tier architecture and describe examples of N-tier architecture & 521 & 3.18 & 1.12 \\
28. I can describe and utilize the core ideas of object-oriented programming when designing software systems & 521 & 3.99 & 0.89 \\
29. I can analyze design trade-offs (e.g., performance vs. scalability) when selecting a software architecture & 521 & 3.75 & 0.91 \\
30. I can decompose a complex software system into components and modules that promote maintainability and reuse & 521 & 3.76 & 0.92 \\
31. I can apply the pipe-and-filter architectural style to design systems that process and transform data through sequential stages & 521 & 3.18 & 1.11 \\
32. I can justify architectural design decisions by documenting the rationale and the attributes they address & 521 & 3.57 & 0.98 \\
33. I can design a blackboard architecture for systems that require shared knowledge or iterative problem-solving among components & 521 & 3.34 & 1.10 \\
34. I can communicate effectively (both orally and in writing) with different teammates. & 522 & 4.40 & 0.79 \\
35. I can choose an appropriate channel for formal (e.g., official emails) or informal (e.g., texting) communication & 522 & 4.44 & 0.75 \\
36. I can collaboratively plan user stories to organize project responsibilities and coordinate coding tasks. & 522 & 4.23 & 0.80 \\
37. I can use GitHub for version control and collaborative software development & 522 & 4.38 & 0.75 \\
38. I can manage branches, commits, and pull requests on GitHub & 522 & 4.29 & 0.85 \\
39. I can use issues, comments, and notifications on GitHub for coordination & 522 & 4.22 & 0.87 \\
40. I can troubleshoot Git-related problems (e.g., merge conflicts) in team settings & 522 & 4.06 & 0.93 \\
41. I can take the initiative to organize meetings, progress checks, and help my teammates meet deadlines & 522 & 4.30 & 0.81 \\
42. I can mediate disagreements among team members & 522 & 4.19 & 0.85 \\
43. I can complete my assigned responsibilities on time and communicate any delays to my team & 522 & 4.39 & 0.74 \\
44. I can review my teammates' code and provide constructive, respectful suggestions to improve their work & 522 & 4.15 & 0.82 \\
45. I can collaborate effectively with others to resolve merge or integration conflicts & 522 & 4.23 & 0.82 \\
46. I can present and pitch ideas for professional software development environments & 522 & 4.15 & 0.87 \\
47. I can support my teammates when they face technical (e.g., runtime errors) or organizational (e.g., conflicts with other team members) challenges. & 522 & 4.24 & 0.74 \\
48. I can contribute to building trust and maintaining positive relationships within my team. & 522 & 4.39 & 0.76 \\
49. I can plan and estimate the effort and time required for software development tasks. & 522 & 3.94 & 0.85 \\
50. I can break down requirements into modules and software development tasks. & 522 & 4.08 & 0.76 \\
51. I can design a system architecture that satisfies the given requirements. & 522 & 3.89 & 0.80 \\
52. I can write clean, well-documented, and maintainable code that complies with team standards. & 522 & 4.06 & 0.75 \\
53. I can prepare and execute deployment procedures (i.e., releasing software components into production). & 522 & 3.82 & 0.92 \\
54. I can adapt existing code to accommodate changing requirements. & 522 & 4.06 & 0.78 \\
55. I can find missing or conflicting parts in requirement specifications. & 522 & 3.98 & 0.83 \\
56. I can model requirements using appropriate diagrams (e.g., entity-relationship) that clearly show system characteristics and behaviors. & 522 & 3.77 & 0.92 \\
57. I can identify key objects and relationships from the requirement documentation. & 522 & 4.02 & 0.80 \\
58. I can translate complex requirements into clear, usable documentation. & 522 & 3.97 & 0.81 \\
59. I can gather user and stakeholder needs through interviews or workshops. & 522 & 3.89 & 0.91 \\
60. I can write use cases and user stories that reflect real user needs. & 522 & 4.04 & 0.82 \\
61. I can prioritize requirements based on user and project goals. & 522 & 4.13 & 0.78 \\
62. I can update requirement documents when project conditions change. & 522 & 4.15 & 0.75 \\
63. I can connect requirements to code, design, and testing artifacts. & 522 & 4.08 & 0.77 \\
64. I can check that implemented features meet stakeholder expectations. & 522 & 4.07 & 0.80 \\
65. I can manage requirement changes using documentation or version control tools. & 522 & 4.07 & 0.77 \\
66. I can negotiate priorities and trade-offs with relevant stakeholders. & 522 & 3.84 & 0.94 \\
67. I can keep requirements traceable throughout the project lifecycle. & 522 & 3.98 & 0.82 \\
68. I can recognize and address code smells or design flaws in different programming environments & 520 & 3.76 & 0.92 \\
69. I can refactor existing code to enhance its structure and maintainability across various software systems & 520 & 3.85 & 0.88 \\
70. I can utilize quality analysis tools to assess and improve code health across various software systems & 519 & 3.69 & 0.98 \\
71. I can create effective test cases from specifications or existing software across different contexts & 519 & 3.85 & 0.87 \\
72. I can automate and execute tests to verify software behavior across multiple software environments & 519 & 3.77 & 0.96 \\
73. I can apply standard coding, documentation, and process guidelines consistently in diverse settings & 520 & 3.99 & 0.83 \\
74. I can conduct and contribute effectively to peer code reviews with my teammates & 520 & 4.02 & 0.80 \\
75. I can use software engineering metrics to evaluate and monitor quality across multiple projects & 520 & 3.80 & 0.90 \\
76. I can uphold ethical and professional standards when working on software projects & 520 & 4.22 & 0.77 \\
77. I can identify and prioritize risks that impact software quality across different contexts & 519 & 3.96 & 0.82 \\
78. I can use version control, Continuous Integration/Continuous Delivery, or other tools to maintain continuous quality over time. & 519 & 3.94 & 0.93 \\
79. I can apply lessons learned from previous projects to enhance quality in future work. & 519 & 4.23 & 0.76 \\
80. I can detect and mitigate potential security vulnerabilities across different platforms and technologies. & 519 & 3.64 & 0.99 \\
81. I can implement techniques to improve reliability and fault tolerance under different conditions. & 519 & 3.77 & 0.89 \\
82. I can evaluate software performance and stability across diverse usage environments. & 519 & 3.75 & 0.91 \\
83. I can identify and refactor code to improve its quality and maintainability. & 519 & 3.93 & 0.88 \\
84. I can design software components that protect users' data privacy and security. & 519 & 3.66 & 1.02 \\
85. I can select and manage appropriate cloud-based platforms (SaaS, PaaS, IaaS) for software deployment. & 519 & 3.39 & 1.15 \\
86. I can apply ethical principles when making software design decisions. & 519 & 4.18 & 0.78 \\
87. I can utilize Development and Operations (DevOps) practices, such as continuous integration and automated testing, to enhance software quality. & 519 & 3.62 & 1.03 \\
\end{longtable}
\endgroup

\end{document}